\documentclass[conference]{IEEEtran}
\IEEEoverridecommandlockouts
\usepackage{cite}
\usepackage{amsmath,amssymb,amsthm,amsfonts,multicol, nccmath}
\newcommand\Tau{\mathcal{T}}
\usepackage{algorithm, algpseudocode}
\usepackage{graphicx}
\usepackage{textcomp}
\usepackage[hidelinks, draft]{hyperref}
\usepackage{balance}
\usepackage{xcolor}
\usepackage{subcaption}
\usepackage{float}
\def\BibTeX{{\rm B\kern-.05em{\sc i\kern-.025em b}\kern-.08em
		T\kern-.1667em\lower.7ex\hbox{E}\kern-.125emX}}

\begin{document}
	
	\title{On the Semi-Blind Mutually Referenced Equalizers for MIMO Systems\\
    	\thanks{This work was supported by the National Foundation for Science and Technology Development of Vietnam under Grant 01/2019/TN. Correspondence: Nguyen Linh Trung (linhtrung@vnu.edu.vn).}
	}
	
\makeatletter
\newcommand{\linebreakand}{%
\end{@IEEEauthorhalign}
\hfill\mbox{}\par
\mbox{}\hfill\begin{@IEEEauthorhalign}
}
\makeatother
	
\author{
    \IEEEauthorblockN{Do Hai Son\IEEEauthorrefmark{2},
    Karim Abed-Meraim\IEEEauthorrefmark{3},
    Tran Trong Duy\IEEEauthorrefmark{2}, 
    Nguyen Linh Trung\IEEEauthorrefmark{2},
    Tran Thi Thuy Quynh\IEEEauthorrefmark{2}}
    \linebreakand
    \IEEEauthorrefmark{2} AVITECH, VNU University of Engineering and Technology, Hanoi, Vietnam\\
    \IEEEauthorrefmark{3} PRISME Laboratory (IUF member), University of Orl\'eans, Orl\'eans, France\\
}

\maketitle

\begin{abstract}
	Minimizing training overhead in channel estimation is a crucial challenge in wireless communication systems. This paper presents an extension of the traditional blind algorithm, called ``Mutually referenced equalizers'' (MRE), specifically designed for MIMO systems. Additionally, we propose a novel semi-blind method, SB-MRE, which combines the benefits of pilot-based and MRE approaches to achieve enhanced performance while utilizing a reduced number of pilot symbols. Moreover, the SB-MRE algorithm helps to minimize complexity and training overhead and to remove the ambiguities inherent to blind processing. The simulation results demonstrated that SB-MRE outperforms other linear algorithms, i.e., MMSE, ZF, and MRE, in terms of training overhead symbols and complexity, thereby offering a promising solution to address the challenge of minimizing training overhead in channel estimation for wireless communication systems.
	  
\end{abstract}


\section{Introduction}\label{Intro}

Multi-Input Multi-Output (MIMO) communication systems play a vital role in achieving reliable high data rate transmission while improving spectrum efficiency and channel capacity~\cite{George2017}. In order to mitigate channel distortions and separate source signals, MIMO systems employ an increased number of training sequences, usually referred to as pilots, which do not carry any useful data. However, this leads to a decrease in spectrum efficiency. In the 1990s, many studies proposed `blind' (B) approach algorithms~\cite{abed1997} to estimate the wireless channel without any pilot symbol. However, blind methods often have high complexity or require statistical information that is not ready for real-world systems. After all, pilot-based channel estimation techniques have dominated blind methods up to now. 

In recent years, there has been a growing interest in the combination of pilot-based and blind methods, known as `semi-blind' (SB) methods, which aim to reduce the number of pilots while maintaining stable performance~\cite{Ladaycia2019}. In this work, we focus on the blind ``Mutually referenced equalizers'' (MRE)~\cite{original} algorithm, which is fast, global convergence, and flexible in implementation (i.e., Batch, LMS, RLS,~\ldots). Since 1997, several studies have been proposed to develop the MRE algorithm.
\cite{GesbertSPAWC} firstly proposed recursive least-squares (RLS) implementation for the MRE method. In \cite{Gesbert1997}, Gesbert \textit{et al.} introduced the MIMO version of the MRE method. In 2000, J.~van der Veen \textit{et al.}~\cite{Veen2000} proposed the combination of MRE with another blind algorithm, which is ``Constant modulus'' (CMA) for SIMO systems.~\cite{Yu2015} also used MRE method for SIMO models but reduced the complexity by computing only $2$ instead of $K$ equalizers as the original paper. Generally, most of the studies reviewed above and others tried to develop the MRE algorithm in the blind approach. As explained above, they often have one or more drawbacks, e.g., lack of support for MIMO; the need for intensive computational resources when the number of equalizers is large; and the requirement for several pieces of channel information, i.e., channel order and delay between receivers.

This motivates us to propose the SB-MRE method, a semi-blind approach for the MRE algorithm in MIMO systems. In detail, a few pilot symbols are used to improve the performance of the B-MRE component. To further enhance its effectiveness, we aim to reduce the overall cost of the SB-MRE method through two key improvements. Firstly, we decrease the complexity of the B-MRE component by reducing the number of equalizers to just 2. Secondly, we employ a straightforward adaptive algorithm that effectively minimizes the number of pilot symbols required in the SB-MRE method.

Our contribution in this paper is to propose an effective SB-MRE method for MIMO systems. The system model and B-MRE algorithm for MIMO systems are presented in section~\ref{review}. The novel SB-MRE algorithm is shown in section~\ref{Method}. Two methods to reduce the cost of SB-MRE are presented in section~\ref{cost}. At last,  we conduct numerical experiments to compare the performance of SB-MRE with other linear estimation methods.

\begin{fleqn}
    \begin{equation*}
        \begin{array}{lll}
            \textit{Notations:}& \mathbf{X}^\top  & \text{Transpose matrix of } \mathbf{X}. \\
            & \mathbf{X}^H     & \text{Hermitian matrix of } \mathbf{X}. \\
            & \mathbf{I}_K     & \text{Indentity matrix of shape } K. \\
            & \otimes & \text{Kronecker product operator.} \\
            & \mathbb{E}()     & \text{Statistical expectation.} \\
            & \mathbf{0} & \text{Matrix of zeros.} 
        \end{array}
    \end{equation*}
\end{fleqn}

\section{MIMO-MRE}\label{review}
This section introduces the mathematical model of MIMO wireless communications used in this work. After that, the MRE algorithm for the MIMO model is briefly reviewed.

\subsection{System model} \label{sm}
\begin{figure}[ht]
    \centering
    \includegraphics[width=\linewidth]{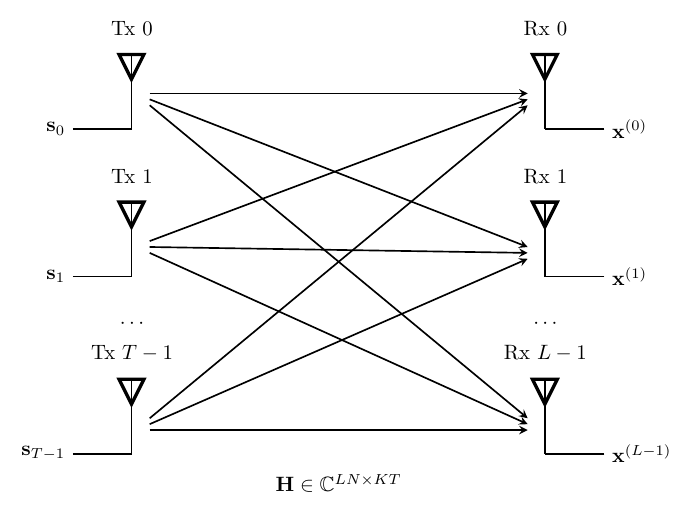}
    \caption{Conventional MIMO communication model.}
    \label{fig:sys_model}
\end{figure}

The MIMO model, illustrated in Fig.~\ref{fig:sys_model}, is composed of $T$~transmitters and $L$~receivers. Each channel between $t$-th transmitter and $l$-th receiver is formulated as a $M+1$ coefficients vector. At a time, $N$ received symbols are simultaneously captured on each receiver. At time $n$, the following equation expresses the system model.
\begin{equation}
    \mathbf{x}(n) = \sum_{t=0}^{T-1}\mathbf{H}_t \mathbf{s}_t(n) + \mathbf{w}_t,
\end{equation}
where $\mathbf{s}_t(n) \in \mathbb{C}^{M+N \times 1}$ is the transmit symbols from the $t$-th transmitter. $\mathbf{H}_t$ is the channel convolution matrix between $t$-th transmitter and $L$ receivers. $\mathbf{H}_t \in \mathbb{C}^{LN \times K}$ is assumed to be a full column rank ($K = M+N$) matrix. $\mathbf{x}(n) \in \mathbb{C}^{LN \times 1}$ denotes the observed signals and $\mathbf{w}_t \in \mathbb{C}^{LN \times 1}$ stands for additive white Gaussian noise matrix. Assume that channel and additive noise between each channel are i.i.d and have distributed $\mathcal{CN}(0, \sigma_{\mathbf{H}_t}^2 \mathbf{I})$ and $\mathcal{CN}(0, \sigma^2 \mathbf{I})$, respectively.

\begin{equation*}
    \mathbf{s}_t(n) = [s_t(n), s_t(n-1),\ldots,s_t(n-K+1)]^\top,
\end{equation*}

\begin{equation*}
    \small
    \mathbf{H}_t \hspace{-0.1cm} = \hspace{-0.7cm}
    \begin{array}{cc}
         & \underset{\longleftrightarrow}{K}\\
         & \left(\begin{array}{cccccc}
    h_{t,0}^{(0)} & \cdots & h_{t,M}^{(0)} & 0 & \cdots & 0 \\
    \vdots & \cdots & \ddots & \cdots & \ddots & 0 \\
    0 & \cdots & 0 & h_{t,0}^{(0)} & \cdots & h_{t,M}^{(0)} \\
    \vdots & \cdots & \vdots & \cdots & \cdots & \vdots \\
    h_{t,0}^{(L-1)} & \cdots & h_{t,M}^{(L-1)} & 0 & \cdots & 0 \\
    \vdots & \cdots & \ddots & \cdots & \ddots & 0 \\
    0 & \cdots & 0 & h_{t,0}^{(L-1)} & \cdots & h_{t,M}^{(L-1)}
    \end{array}
    \right) 
    \end{array}
    \Big\updownarrow LN,
\end{equation*}

\begin{fleqn}
    \begin{equation*}
        \begin{aligned}
            \mathbf{x}(n) = \big[x^{(0)}(n), \cdots, &x^{(0)}(n-N+1), \cdots, 
            \\ &x^{(L-1)}(n), \cdots, x^{(L-1)}(n-N+1)\big]^\top.
        \end{aligned}
    \end{equation*} 
\end{fleqn}

\subsection{Brief review of MRE method}

Generally, MRE uses an $N$-taps linear equalizer to filter each channel. Let $\mathbf{g}_{t, i}  \in \mathbb{C}^{LN \times 1}$ be an $i$-delay equalizer and $t$-th transmitter. For each transmitter, the number of equalizers equals the maximum delay, which is $K$. For \mbox{$i=0, \ldots, K-1$}, at time $n$, we have
\begin{equation}
    \mathbf{g}^H_{t, i} * \mathbf{x}(n)=\sum_{l=0}^{L-1}\sum_{k=0}^{N-1} g^H_{t,i}(k) x^{(l)}(n-k) \approx \mathbf{s}_t(n-i),
\end{equation}
\begin{equation}
\begin{aligned}
    \mathbf{g}_{t,i}=\big[g_{t, i}^{(0)}(0), \ldots, &g_{t, i}^{(0)}(N - 1), \ldots, \\
    &g_{t, i}^{(L-1)}(0), \ldots, g_{t, i}^{(L-1)} (N-1) \big]^\top.
\end{aligned}
\end{equation}
The equalizers matrix for $t$-th transmitter is $\mathbf{G}_t \in \mathbb{C}^{LN \times K}$ as follows:
\begin{equation}
    \mathbf{G}_t = [\mathbf{g}_{t, 0}, \ldots, \mathbf{g}_{t, K-1}].
\end{equation}

In the noise-free case, the transmitted symbols can be perfectly recovered with $\bar{\mathbf{G}}$ is any left inverse of $\mathbf{H}$ since
\begin{equation}
    \begin{aligned}
        \relax[\mathbf{G}_0, \ldots, \mathbf{G}_{T-1}]^{H} 
        \mathbf{x}(n)
        &= [\mathbf{s}_0^\top(n), \ldots, \mathbf{s}^\top_{T-1}(n)]^\top, \\
        \bar{\mathbf{G}}^H \mathbf{x}(n) &= \bar{\mathbf{s}}(n).
    \end{aligned}   
\end{equation}

In noisy case, to estimate $\bar{\mathbf{G}}$, the MRE method exploits the delay diversity from multi-channel, $\mathbf{g}_i^H \mathbf{x}(n) = \mathbf{g}_{i+1}^H \mathbf{x}(n+1)$, to determine the full set of channel inverses. Where $\mathbf{g}$ is the vector form of $\bar{\mathbf{G}}$ equalizers matrix as shown in Eq.~(\ref{eq:vecG}). The unconstrained MRE cost function of $\bar{\mathbf{G}}$ is given by:
\begin{equation}
    \mathcal{J}(\bar{\mathbf{G}})=\mathbf{g}^H \mathcal{R} \mathbf{g},
\end{equation}
where $\mathcal{R} \in \mathbb{C}^{LNKT \times LNKT}$ is the matrix of $\mathbf{x}(n)$ and $\mathbf{x}(n+1)$ observed signals, which is given by:
\begin{equation}
\label{eq:R}
\mathcal{R} \stackrel{\text { def }}{=} \mathbb{E}\left(\mathbf{U}^{H} \mathbf{U}\right),
\end{equation}
with
\begin{equation}
\label{eq:U}
\mathbf{U} = \left(\mathbf{I}_{T (K-1)}, \mathbf{0}\right) \otimes \mathbf{x}^{H}(n)-\left(\mathbf{0}, \mathbf{I}_{T (K-1)}\right) \otimes \mathbf{x}^{H}(n+1).
\end{equation}
Under the quadratic constraint~\cite{original}, the unique stable minimum of $\mathbf{g}$ is estimated by selecting the smallest eigenvector of $\mathcal{R}$.



\section{Propose SB-MRE}\label{Method}

In $t$-th transmitter, a block data $\mathbf{s}_t$ is considered to send, including $N_p$ pilot symbols and $N_s - N_p$ data symbols.
\begin{equation}
\begin{aligned}
    \mathbf{s}_t(n) = [s_t(n), \ldots, &s_t\left(n-N_{p} + 1\right), \\
    &s_t\left(n - N_p\right), \ldots, s_t\left(n - N_s+1\right)]^\top.
\end{aligned}
\end{equation}

Pilot signals estimate the full set of channel inverse by the least-square method.
\begin{equation}
    \label{eq:ls}
    \hat{\mathbf{G}} = \arg \underset{\bar{\mathbf{G}} \in \mathbb{C}^{LN \times KT}}{\min} \sum_{i=N-1}^{N_{p} - 1}\|\bar{\mathbf{s}}(n)- \bar{\mathbf{G}}^H \mathbf{x}(n)\|_F^2 .
\end{equation}

The combining of pilot-based and blind MRE is a constrained optimization that can readily solve by the Lagrange multiplier method~\cite{bertsekas2014constrained}. The total cost function of SB-MRE will be
\begin{equation}
\label{eq:cost}
    \mathcal{J}(\bar{\mathbf{G}})=\sum_{i=N-1}^{N_{p} - 1}\|\bar{\mathbf{s}}(n)- \bar{\mathbf{G}}^H \mathbf{x}(n)\|_F^2 +\lambda \mathbf{g}^H \mathcal{R} \mathbf{g},
\end{equation}
with $\lambda$ is a weighting factor, $\mathcal{R}$ in the quadratic form of the blind MRE criterion as shown in Eq.~(\ref{eq:R}), and $\mathbf{g}$ is the vector form of $\bar{\mathbf{G}}$.
\begin{equation}
\label{eq:vecG}
    \begin{aligned}
        \mathbf{g} = \operatorname{vec}(\bar{\mathbf{G}}) &=\left[\vec{\mathbf{G}}_0^\top, \vec{\mathbf{G}}_1^\top, \ldots, \vec{\mathbf{G}}_{K-1}^\top\right]^\top, \\
        \vec{\mathbf{G}}_i &= \left[\begin{array}{ll}
        \mathbf{g}_{0, i}^\top, \mathbf{g}_{1, i}^\top, \ldots, \mathbf{g}_{T-1, i}^\top
        \end{array}\right]^\top.
    \end{aligned}
\end{equation}

Without loss of generality, the least-square expression of Eq.~(\ref{eq:cost}) is conjugate transposed and the sum operator is turned into matrix forms of $\widetilde{\mathbf{S}}$ and $\widetilde{\mathbf{X}}$. The cost function is expressed as follows:
\begin{equation}
    \begin{aligned}
    \mathcal{J}(\bar{\mathbf{G}})&=\sum_{i=N-1}^{N_{p} - 1}\left\|{\bar{\mathbf{s}}^H(n)}-\mathbf{x}^H(n) \bar{\mathbf{G}}\right\|^2_F +\lambda \mathbf{g}^H \mathcal{R} \mathbf{g}\\
    &=\left\|\widetilde{\mathbf{S}}^H-\widetilde{\mathbf{X}}^H \bar{\mathbf{G}}\right\|^2_F +\lambda \mathbf{g}^H \mathcal{R} \mathbf{g}.
    \end{aligned}
\end{equation}
where $\widetilde{\mathbf{S}}, \widetilde{\mathbf{X}}$ are the matrices of shape $\mathbb{C}^{KT \times (N_p -N +1)}$ and $ \mathbb{C}^{LN \times (N_p-N+1)}$, respectively.
\begin{equation*}
    \begin{aligned}
        \widetilde{\mathbf{S}} &= [\bar{\mathbf{s}}(N-1), \ldots, \bar{\mathbf{s}}\left(N_{p} - 1\right)], \\
        \widetilde{\mathbf{X}} &= [\mathbf{x}(N-1), \ldots, \mathbf{x}\left(N_{p} - 1\right)].
    \end{aligned}
\end{equation*}

The least-square expression is vectorized and thanks to the property for vector, i.e., $\operatorname{vec}(\mathbf{AXB}) = (\mathbf{B}^\top \otimes \mathbf{A}) * \operatorname{vec}(\mathbf{X})$. The SB-MRE cost function turned into
\begin{equation}
\label{eq:cost_final}
    \begin{aligned}
    \mathcal{J}(g) &= \left\|\operatorname{vec}(\widetilde{\mathbf{S}}^H) - (\mathbf{I}_{KT} \otimes \widetilde{\mathbf{X}}^H) \operatorname{vec}(\bar{\mathbf{G}})\right\|^2_F + \lambda \mathbf{g}^H \mathcal{R} \mathbf{g}\\
     &= \left\| \bar{\mathbf{s}} - \mathbf{A} \mathbf{g} \right\|^2_F + \lambda \mathbf{g}^H \mathcal{R} \mathbf{g} \\
     &= \mathbf{g}^H \mathbf{A}^H \mathbf{A} \mathbf{g} + \left\| \bar{\mathbf{s}} \right\|^2_F - 2\Re (\mathbf{g}^H \mathbf{A}^H \bar{\mathbf{s}}) + \lambda \mathbf{g}^H \mathcal{R} \mathbf{g}.  \end{aligned}
\end{equation}

In order to find minimum cost of Eq.~(\ref{eq:cost_final}), let derivative $\mathcal{J}(\mathbf{g})$ with respect to $\mathbf{g}$ as follows:
\begin{equation}
\begin{aligned}
\frac{\partial \mathcal{J}}{\partial \mathbf{g}}(\mathbf{g}) &= 0, \\
\left(\mathbf{A}^H \mathbf{A}+\lambda \mathcal{R}\right) \mathbf{g} &= \mathbf{A}^H \bar{\mathbf{s}}.
\end{aligned}
\end{equation}

The final equalizers matrix in vector form of the proposed SB-MRE method is obtained through
\begin{equation}
    \mathbf{g}_{SB}=\left(\mathbf{A}^H \mathbf{A} + \lambda \mathcal{R}\right)^{-1} \mathbf{A}^H \bar{\mathbf{s}}.
\end{equation}

\section{Reduce the cost}\label{cost}
In the ensuing, we aim to reduce the cost of the proposed SB-MRE algorithm by addressing two key factors, i.e., reducing the complexity of the B-MRE component and minimizing the training overhead.

\subsection{Reducing the complexity of the B-MRE part}
In the original study, the overall complexity of the blind MRE method is $\mathcal{O}(LNKT)$~\cite{original}. Although all $K$ equalizers are estimated for each transmitter, only one is ultimately utilized. However, this computational burden becomes unnecessary as $N$ increases. Hence, in this section, we first considerably reduce the number of equalizers to $2$, i.e., the $0$-th and \mbox{$(K-1)$-th} equalizers. As a result, the overall complexity is reduced to $\mathcal{O}(LNT)$ and the equalizer matrix for the $t$-th transmitter can be represented as follows:
\begin{equation}
    \mathbf{V}_{t} = [\mathbf{g}_{t, 0}, \; \mathbf{g}_{t, K-1}].
\end{equation}
Followed by the estimated signal source of $t$-th transmitter will be
\begin{equation}
    \mathbf{V}_t^H \mathbf{x}(n) = [s_t(n), s_t(n-K+1)]^\top = \mathbf{s}_{t}(n).
\end{equation}
Following that, we do not have to compute the full rank of $\mathcal{R}$ as the blind approach. Eq.~(\ref{eq:U}) is modified to
\begin{equation*}
\mathbf{U} = \left(\mathbf{I}_{T}, \mathbf{0}\right) \otimes \mathbf{x}^{H}(n)-\left(\mathbf{0}, \mathbf{I}_{T}\right) \otimes \mathbf{x}^{H}(n+K-1).
\end{equation*}
\subsection{Reducing the training overhead for SB-MRE}

In the least-square method, as indicated in Eq.~(\ref{eq:ls}), the performance of SB-MRE is primarily influenced by the number of pilot symbols ($N_p$). However, increasing the number of pilot symbols results in a decrease in spectral efficiency. To address this issue, we propose an adaptive algorithm that determines the minimum number of pilots required for the SB-MRE method. 

First, we establish an assumption that after a few transmission sessions, the user equipment (UE) provides feedback to the base station (BTS) regarding the average symbol error rate (SER). This assumption is inspired by the feedback of the block error rate (BLER) in 5G standards~\cite{r16}. Second, taking into account the channel state information (CSI), UE characteristics, and the specific service requirements, a target SER value ($\Tau$) is determined for a given transmission period. Subsequently, we define a straightforward loss function that measures the deviation from the target SER value as follows:
\begin{equation}
    \mathcal{L} (\operatorname{SER}) = \log_{10}(\operatorname{SER}) - \log_{10}(\Tau).
\end{equation}
The updating formula for the number of pilots $N_p$ is
\begin{equation}
    N_p = N_p + \delta * \mathcal{L} (\operatorname{SER}),
\end{equation}
where $\delta$ is the learning rate. \\

Initially, the parameter $\delta$ is set to $1$. If the feedback SER has not yet reached the target SER ($\Tau$), $\delta = \delta_1$. However, once the target SER is satisfied, the number of pilots is reduced to improve data efficiency, and during this phase, $\delta$ remains fixed at $1$. The algorithm~\ref{alg:cap} shows our method to adapt the number of pilot symbols for SB-MRE. This approach can be seen as following a multiplicative-increase additive-decrease algorithm for adjusting the number of pilots. The reverse version, additive-increase multiplicative-decrease (AIMD), has been successfully adopted in networking, particularly the TCP congestion control algorithm~\cite{CHIU19891}. Moreover, this scheme has been observed in nature, employed by biological systems, and even adapted for neural circuits~\cite{Suen2022}.

\begin{algorithm}[ht]
    \caption{Adaptive number of pilot symbols for SB-MRE.}\label{alg:cap}
    \hspace*{\algorithmicindent} \textbf{Input:} $\Tau, \delta_1, \operatorname{SER}$ \\
    \hspace*{\algorithmicindent} \textbf{Output: $N_p$} 
    \begin{algorithmic}[1]
        \State $N_p \leftarrow N$
        \While {true}
            \State $\operatorname{SER} = \operatorname{SB-MRE}(N_p)$ 
            \State $\mathcal{L} (\operatorname{SER}) = \log_{10}(\operatorname{SER}) - \log_{10}(\Tau)$
            \If{$\mathcal{L} (\operatorname{SER}) < 0$}
                \State $\delta \quad = 1$
                
            \Else 
                \State $\delta \quad= \delta_1$

            \EndIf
            \State $N_p = N_p + \delta * \mathcal{L} (\operatorname{SER})$
        \EndWhile
    \end{algorithmic}
\end{algorithm}

\section{Simulation Results}\label{SR}
In this section, we present the experimental analysis of the proposed SB-MRE method using the simulation parameters outlined in Table~\ref{tab:simulation_param}. The simulation results are based on an average of 500,000 runs. We begin by comparing the performance of the SB-MRE method against traditional channel estimation algorithms, namely Zero Forcing (ZF) and Minimum Mean Square Error (MMSE)~\cite{Jiang2011}, in terms of SER.

Fig.~\ref{fig:performance} shows that ZF and MMSE outperform the proposed SB-MRE at lower signal-to-noise ratio (SNR) values. This can be attributed to the fact that the influence of the B-MRE component is negligible at low SNR levels. However, as the SNR increases, the proposed SB-MRE gradually catches up to the SER of ZF and MMSE, eventually surpassing them at $\text{SNR}=10\text{dB}$. It is important to note that the proposed SB-MRE algorithm only utilizes 32/256 symbols for pilots, whereas ZF and MMSE require complete knowledge of the CSI. Furthermore, even after reducing the cost of the B-MRE component, the modified SB-MRE (SB-MRE\_rc) still outperforms both the original B-MRE and B-MRE\_rc in terms of SER. This highlights the effectiveness of the proposed modifications in enhancing the performance of the SB-MRE algorithm.
\begin{table}[ht]
\centering
\caption{Simulation parameters}
\label{tab:simulation_param}
    \begin{tabular}{p{3.5cm} | p{4cm}}
    \hline
    \hline
    \textbf{Parameters} & \textbf{Specifications}  \\ \hline
    MIMO                            & $T = 2, L= 4$      \\ \hline
     Modulation                     & QPSK       \\ \hline
    Channel order                   & $M = 3$      \\ \hline
    Windows size                    & $N = 10$     \\ \hline
    Sample size                     & $N_s = 256$  \\ \hline
    Pilots                          & $N_p = 32$   \\ \hline
    Number of blind equalizers      & $2$ \\ \hline
    Weighting factor                & $\lambda = 0.1$   \\ \hline
    Increase rate of $N_p$          & $\delta_1 = 2$ \\ \hline
    Target SER                      & $\Tau = 10^{-4}, \Tau\_\text{rc} = 10^{-2}$ \\ \hline
    \end{tabular}
\end{table}

\begin{figure}[h]
    \centering
    \includegraphics[width=\linewidth]{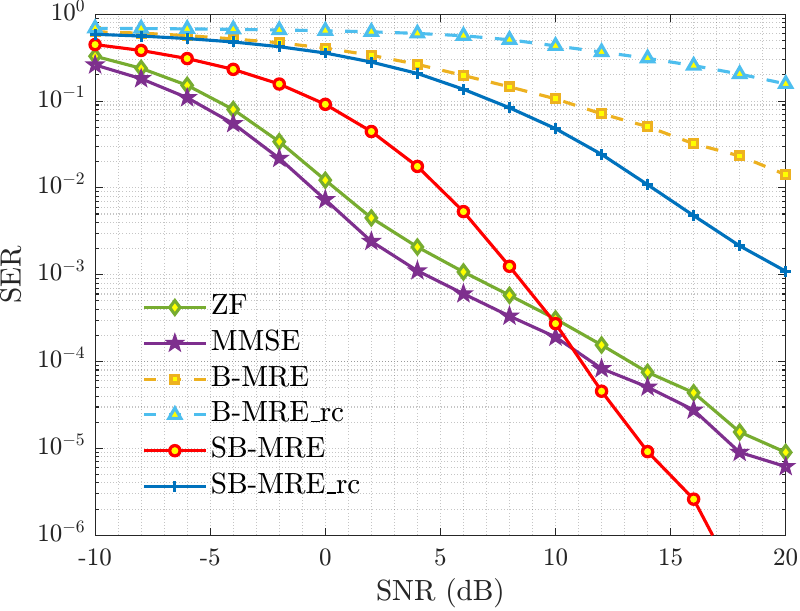}
    \caption{Performance of proposed SB-MRE versus other channel estimation algorithms.}
    \label{fig:performance}
\end{figure}

After that, we simulate to verify the performance of the proposed SB-MRE in different numbers of pilots ($N_p$) and SNR values. As shown in Fig.~\ref{fig:vary_N_p}, $N_p$ and SNR are turned in the range of $[10 \;\; 64]$ pilot symbols and $[5, 10, 15]$~dB, respectively. Overall, SER curves of both SB-MRE and SB-MRE\_rc exhibit a gradual decrease as the number of pilot symbols ($N_p$) and SNR increase. This behavior represents a trade-off between spectrum efficiency and the accuracy of the channel estimation algorithm. At $\text{SNR}=15~\text{dB}$, SB-MRE with $N_p > 40$ archives to SER $\approx 10^{-6}$. In the case of SB-MRE\_rc, when the number of pilot symbols ($N_p$) is increased within the range of $10$ to $40$, there is a clear improvement in the SER curves. The SER decreases noticeably with an increasing number of pilots, indicating better performance. However, once $N_p$ exceeds $40$, the SER curves reach a stable state, and there is little to no further reduction in SER.
\begin{figure}[h]
    \centering
    \begin{subfigure}[b]{\linewidth}
         \centering
         \includegraphics[width=\linewidth]{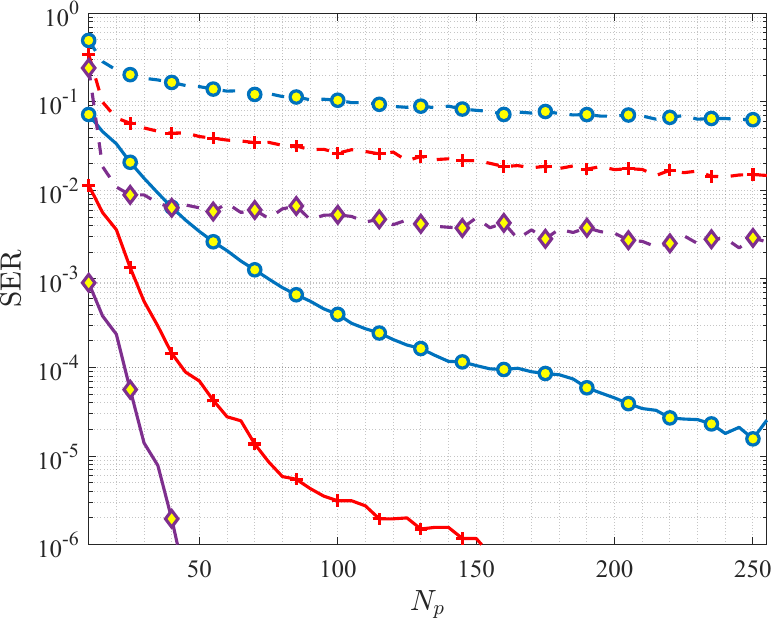}
     \end{subfigure}
     \hfill
     \hspace*{1cm}
     \begin{subfigure}[b]{.8\linewidth}
         \centering
         \includegraphics[width=\linewidth]{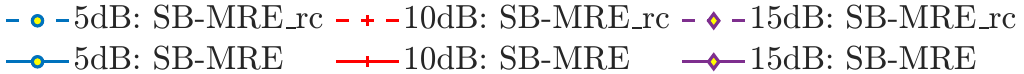}
     \end{subfigure}
     \hfill
    \caption{Performance of proposed SB-MRE under various $N_p$ and SNR values.}
    \label{fig:vary_N_p}
\end{figure}

Fig.~\ref{fig:vary_lambda} demonstrates the impact of the weighting factor ($\lambda$) on the combination of pilot-based and B-MRE. The $\lambda$ is varied within the range of $[0.01 \;\; 0.2]$. For lower SNR levels (i.e., $5$ dB and $10$ dB), we observed a slight reduction in SER curves as $\lambda$ values increased. However, at $\text{SNR}=15~\text{dB}$, the performance of the B-MRE component became prominent, resulting in a significant decrease in the SER curves of SB-MRE. On the other hand, impacts of the weighting factor are negligible in the SB-MRE\_rc version due to the small number of blind equalizers employed.
\begin{figure}[h]
    \centering
    \begin{subfigure}[b]{\linewidth}
         \centering
         \includegraphics[width=\linewidth]{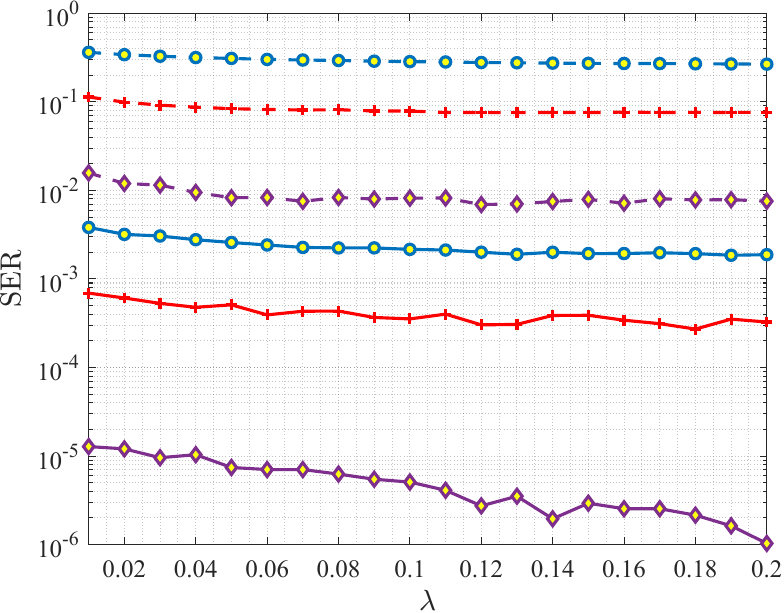}
     \end{subfigure}
     \hfill
     \hspace*{1cm}
     \begin{subfigure}[b]{.8\linewidth}
         \centering
         \includegraphics[width=\linewidth]{fig/legend_2.pdf}
     \end{subfigure}
     \hfill
    \caption{Performance of proposed SB-MRE under various $\lambda$ and SNR values.}
    \label{fig:vary_lambda}
\end{figure}

Finally, the algorithm's performance to reduce the training overhead for SB-MRE is shown in Fig.~\ref{fig:adaptive}. At SNR~$=12$~dB, the weighting factor ($\lambda$) is set at two levels, i.e., $0.1$ and $0.15$. According to simulation results in Fig.~\ref{fig:vary_N_p}, the target SERs of SB-MRE, SB-MRE\_rc are fixed at $10^{-4}$ and $10^{-2}$, respectively. Overall, all scenarios' SER curves converge to their target SER after a certain number of iterations. However, the number of pilot symbols and iterations required for convergence vary significantly. For $\lambda = 0.1$, SB-MRE and SB-MRE\_rc require $51$ and $131$ pilot symbols, along with $20$ and $70$ iterations, respectively. Conversely, for $\lambda = 0.15$, SB-MRE and SB-MRE\_rc only need $13$ and $15$ pilot symbols, along with 1 and 5 iterations, respectively. These results highlight the effectiveness of the adaptive number of pilot symbols method for SB-MRE, with the $\lambda$ value playing a crucial role in enhancing the performance of the SB-MRE algorithm, especially in high SNR scenarios.
\begin{figure}[h]
    \centering
    \begin{subfigure}[b]{\linewidth}
         \centering
         \includegraphics[width=\linewidth]{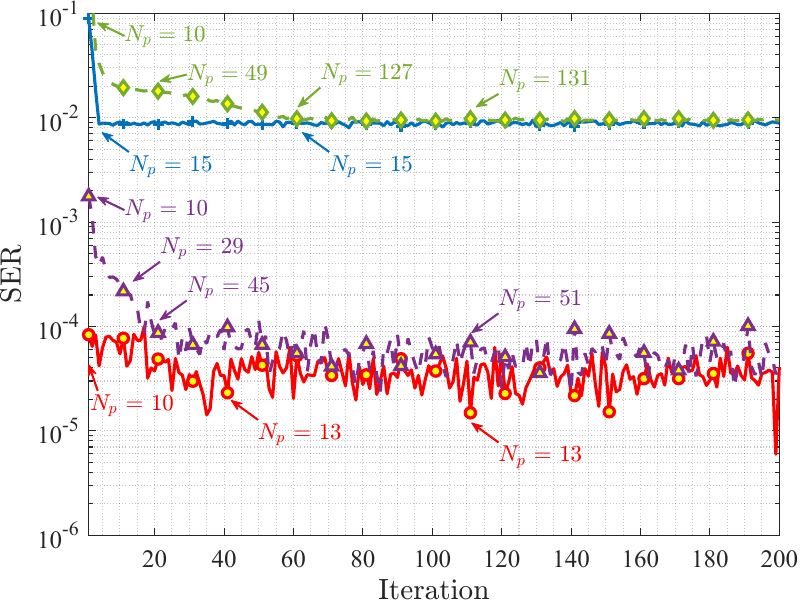}
     \end{subfigure}
     \hfill
     \hspace*{1cm}
     \begin{subfigure}[b]{.8\linewidth}
         \centering
         \includegraphics[width=\linewidth]{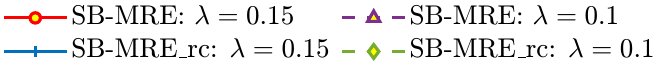}
     \end{subfigure}
     \hfill
    \caption{Adapt the number of pilot symbols method for SB-MRE. The simulation parameters are $\Tau~=~10^{-4}, \Tau\_\text{rc}~=~10^{-2},$ and SNR~$=12$~dB.}
    \label{fig:adaptive}
\end{figure}

\section{Conclusion}

In this paper, we proposed a semi-blind version of the mutually referenced equalizers (MRE) algorithm designed for MIMO systems. Our algorithm uses a small number of pilot symbols to enhance the performance of the B-MRE. Furthermore, the SB-MRE algorithm also reduces the number of blind equalizers and pilot symbols required. Simulation results demonstrate that SB-MRE outperforms other B-MRE approaches and linear methods, particularly in high SNR scenarios. Additionally, we analyze the impact of the blind component and highlight the effectiveness of adapting the number of pilot symbols.

	
\bibliographystyle{IEEEtran}
\bibliography{library.bib}	

\begin{thebibliography}{10}
\providecommand{\url}[1]{#1}
\csname url@samestyle\endcsname
\providecommand{\newblock}{\relax}
\providecommand{\bibinfo}[2]{#2}
\providecommand{\BIBentrySTDinterwordspacing}{\spaceskip=0pt\relax}
\providecommand{\BIBentryALTinterwordstretchfactor}{4}
\providecommand{\BIBentryALTinterwordspacing}{\spaceskip=\fontdimen2\font plus
\BIBentryALTinterwordstretchfactor\fontdimen3\font minus
  \fontdimen4\font\relax}
\providecommand{\BIBforeignlanguage}[2]{{%
\expandafter\ifx\csname l@#1\endcsname\relax
\typeout{** WARNING: IEEEtran.bst: No hyphenation pattern has been}%
\typeout{** loaded for the language `#1'. Using the pattern for}%
\typeout{** the default language instead.}%
\else
\language=\csname l@#1\endcsname
\fi
#2}}
\providecommand{\BIBdecl}{\relax}
\BIBdecl

\bibitem{George2017}
G.~George, R.~K. Mungara, A.~Lozano, and M.~Haenggi, ``Ergodic spectral
  efficiency in mimo cellular networks,'' \emph{IEEE Transactions on Wireless
  Communications}, vol.~16, no.~5, pp. 2835--2849, May 2017.

\bibitem{abed1997}
K.~Abed-Meraim, W.~Qiu, and Y.~Hua, ``Blind system identification,''
  \emph{Proceedings of the IEEE}, vol.~85, no.~8, pp. 1310--1322, Aug. 1997.

\bibitem{Ladaycia2019}
A.~Ladaycia, A.~Belouchrani, K.~Abed-Meraim, and A.~Mokraoui, ``Semi-blind
  mimo-ofdm channel estimation using expectation maximisation like
  techniques,'' \emph{IET Communications}, vol.~13, no.~20, pp. 3452--3462,
  Dec. 2019.

\bibitem{original}
D.~Gesbert, P.~Duhamel, and S.~Mayrargue, ``On-line blind multichannel
  equalization based on mutually referenced filters,'' \emph{IEEE Transactions
  on Signal Processing}, vol.~45, no.~9, pp. 2307--2317, Sept. 1997.

\bibitem{GesbertSPAWC}
D.~Gesbert and P.~Duhamel, ``Unimodal blind adaptive channel equalization: an
  rls implementation of the mutually referenced equalizers,'' in \emph{First
  IEEE Signal Processing Workshop on Signal Processing Advances in Wireless
  Communications}, Paris, France, Apr. 1997, pp. 29--32.

\bibitem{Gesbert1997}
D.~Gesbert, A.~Paulraj, and P.~Duhamel, ``Blind joint multiuser detection using
  second-order statistics and structure information,'' in \emph{Proceedings of
  40th Midwest Symposium on Circuits and Systems}, vol.~2, Sacramento, CA, USA,
  Aug. 1997, pp. 1252--1255.

\bibitem{Veen2000}
A.-J. van~der Veen and A.~Trindade, ``Combining blind equalization with
  constant modulus properties,'' in \emph{34th Asilomar Conference on Signals,
  Systems and Computers}, vol.~2, Pacific Grove, CA, USA, Oct. 2000, pp.
  1568--1572.

\bibitem{Yu2015}
C.~Yu and L.~Xie, ``On recursive blind equalization in sensor networks,''
  \emph{IEEE Transactions on Signal Processing}, vol.~63, no.~3, pp. 662--672,
  Feb. 2015.

\bibitem{bertsekas2014constrained}
D.~P. Bertsekas, \emph{Constrained optimization and Lagrange multiplier
  methods}.\hskip 1em plus 0.5em minus 0.4em\relax Academic press, 2014.

\bibitem{r16}
``Monitor of bler performance,'' ETSI, Standard 3GPP TS 32.425 version 16.2.0
  Release 16, Mar. 2019.

\bibitem{CHIU19891}
D.-M. Chiu and R.~Jain, ``Analysis of the increase and decrease algorithms for
  congestion avoidance in computer networks,'' \emph{Computer Networks and ISDN
  Systems}, vol.~17, no.~1, pp. 1--14, Jun. 1989.

\bibitem{Suen2022}
J.~Y. Suen and S.~Navlakha, ``A feedback control principle common to several
  biological and engineered systems,'' \emph{Journal of The Royal Society
  Interface}, vol.~19, no. 188, p. 20210711, Mar. 2022.

\bibitem{Jiang2011}
Y.~Jiang, M.~K. Varanasi, and J.~Li, ``Performance analysis of zf and mmse
  equalizers for mimo systems: An in-depth study of the high snr regime,''
  \emph{IEEE Transactions on Information Theory}, vol.~57, no.~4, pp.
  2008--2026, Apr. 2011.

\end{thebibliography}
	
\end{document}